%% file: mast.tex
\documentclass[journal,onecolumn,12pt]{IEEEtran}

\renewcommand{\baselinestretch}{2.0}
\usepackage[nomarkers,nolists]{endfloat}

\usepackage[dvips]{graphicx}
\usepackage{amsmath,amssymb,amsfonts}
\usepackage{color}

\newtheorem{defi}{Definition}
\newtheorem{lemma}{Lemma}
\newtheorem{theo}{Theorem}

\title{Two-Bit Message Passing Decoders for LDPC Codes Over the Binary Symmetric Channel}
\author{Lucile Sassatelli, \IEEEmembership{Member, IEEE,} Shashi Kiran Chilappagari, \IEEEmembership{Member, IEEE,} Bane~Vasic, \IEEEmembership{Senior Member, IEEE,} and David Declercq, \IEEEmembership{Member, IEEE}% <-this % stops a space
\thanks{Manuscript received \today. L. Sassatelli is with the EECS department, Massachusetts Institute of Technology, MA 02139, USA (email: lucisass@mit.edu). S. K. Chilappagari and B. Vasic are with the Department of Electrical and Computer Engineering, University of Arizona, Tucson, Arizona, 85721 USA (emails: \{shashic, vasic\}@ece.arizona.edu). D. Declercq is with ETIS laboratoy, ENSEA/UCP/CNRS-UMR8051, Cergy, France (email: declercq@ensea.fr). Part of this work has been submitted to ISIT 2009, and is currently under review.}
}
\begin{document}
\maketitle
\begin{abstract}
In this paper, we consider quantized decoding of LDPC codes on the binary symmetric channel. The binary message passing algorithms, while allowing extremely fast hardware implementation, are not very attractive from the perspective of performance. More complex decoders such as the ones based on belief propagation exhibit superior performance but lead to slower decoders. The approach in this paper is to consider message passing decoders that have larger message alphabet (thereby providing performance improvement) as well as low complexity (thereby ensuring fast decoding).
We propose a class of message-passing decoders whose messages are represented by two bits. The thresholds for various decoders in this class are derived using density evolution. The problem of correcting a fixed number of errors assumes significance in the error floor region. For a specific decoder, the sufficient conditions for correcting all patterns with up to three errors are derived. By comparing these conditions and thresholds to the similar ones when Gallager B decoder is used, we emphasize the advantage of decoding on a higher number of bits, even if the channel observation is still one bit. 
\end{abstract}


\input{sec1}
\input{sec2}
\input{sec4}

\input{sec3}

\input{sec5}
\input{sec6}

\input{Ack}
\input{AppendixA}
\bibliographystyle{IEEEtran}
\bibliography{bibliographie}
\end{document}

%% file: sec1.tex
\section{Introduction}

The performance of various hard decision algorithms for decoding low-density parity-check (LDPC) codes on the binary symmetric channel (BSC), has been studied in great detail.  Gallager \cite{Gallager} proposed two binary message passing algorithms, namely Gallager A and Gallager B algorithms. A code of length $n$ is said to be $(n,\gamma,\rho)$ regular if all the columns and all the rows of the parity-check matrix of the code have exactly $\gamma$ and $\rho$ non-zero values, respectively. Gallager showed \cite{Gallager} that there exist $(n,\gamma,\rho)$, $\rho>\gamma\geq 3$ regular LDPC codes, with column weight $\gamma$ and row weight $\rho$, for which the bit error probability approaches zero when we operate below the threshold (precise definition will be given in Section \ref{sec_asymp}). Richardson and Urbanke \cite{P8} analyzed ensembles of codes under various message passing algorithms. They also described \textit{density evolution}, a deterministic algorithm to compute thresholds. Bazzi \textit{et al.} \cite{S3} determined exact thresholds for the Gallager A algorithm and outlined methods to analytically determine thresholds of more complex decoders. Zyablov and Pinsker \cite{S4} were the first to analyze LDPC codes under the parallel bit flipping algorithm, and showed that almost all codes in the regular ensemble with $\gamma\geq 5$ can correct a linear number of errors in the code length. Sipser and Spielman \cite{S5} established similar results using expander graph based arguments. Burshtein and Miller \cite{S6} considered expansion arguments to show that message passing algorithms are also capable of correcting a linear number of errors in the code length.

In this paper, we consider hard decision decoding of a fixed LDPC code on the BSC. The BSC serves a useful channel model in applications where there is no access to soft information and also where decoding speed is a major factor. The binary message passing algorithms, while allowing extremely fast hardware implementation, are not very attractive from the perspective of performance. More complex decoders such as the ones based on belief propagation exhibit superior performance but lead to slower decoders. The approach in this paper is to consider message passing decoders that have larger message alphabet (thereby providing performance improvement) as well as low complexity (thereby ensuring fast decoding).

When an LDPC code is decoded by message passing algorithms, the frame error rate (FER) curve has two regions: as the crossover probability $\alpha$ decreases, the slope of the FER curve first increases (the waterfall region), and then sharply decreases. This region of low slope for small $\alpha$ is called the error floor region. The problem of correcting a fixed number of errors assumes significance in the error floor region, where the slope of the FER curve is determined by the weight of the smallest error pattern uncorrectable by the decoder \cite{S8}. 

For iterative decoding over the binary erasure channel (BEC), it is known that avoiding stopping sets \cite{S9} up to size $t$ in the Tanner graph \cite{Tanner81} of the code guarantees recovery from $t$ or less erasures. A similar result for decoding over the BSC is still unknown. The problem of guaranteed error correction capability is known to be difficult and in this paper, we present a first step toward such result by investigating the conditions sufficient to guarantee the correction of three errors in column-weight-four codes.

Column-weight-four codes are of special importance because, under a fixed rate constraint (which implies some fixed ratio of the left and right degrees), the performance of regular LDPC codes under iterative decoding typically improves when the right and left degrees decrease. Burshtein \cite{S7} showed that regular codes with $\gamma=4$, like codes with $\gamma\geq 5$, are capable of correcting a fraction of errors under the parallel bit flipping algorithm. These results are perhaps the best (up to a constant factor) one can hope for in the asymptotic sense. The proofs are, however, not constructive and the arguments cannot be applied for codes of practical length. Chilappagari \textit{et al.} \cite{S11} have shown that for a given column weight, the number of variable nodes having expansion required by the bit flipping algorithm grows exponentially with the girth of the Tanner graph of the code. However, since girth grows only logarithmically with the code length, construction of high rate codes, with lengths in the order of couple of thousands, even with girth eight is difficult.

Generally, increasing the number of correctable errors can be achieved by two methods: (a) by increasing the strength and complexity of a decoding algorithm or/and (b) by carefully designing the code, i.e., by avoiding certain harmful configurations in the Tanner graph. Powerful decoding algorithms such as belief propagation, can correct error patterns which are uncorrectable by simpler binary message passing algorithms like the Gallager A/B algorithm. However, the analysis of such decoders is complicated due to the statistical dependence of messages in finite graphs. It also depends on implementation issues such as the numerical precision of messages. For Gallager B decoder, avoiding certain structures (known as trapping sets \cite{S12}) in the Tanner graph has shown to guarantee the correction of three errors in column-weight-three
codes \cite{S13}, and this paper is an extension of this result.

In this paper, we apply a combination of the above methods to column-weight-four codes. Specifically, we make the following contributions: (a) We propose a class of message-passing decoders whose messages are represented by two bits. We refer to these decoders as to two-bit decoders. (b) For a specific two-bit decoder, we derive sufficient conditions for a code with Tanner graph of girth six to correct three errors.

The idea of using message alphabets with more than two values for the BSC was first proposed by Richardson and Urbanke in \cite{P8}. They proposed a decoder with erasures in the message alphabet. The messages in such a decoder hence have three possible values. They showed that such decoders exhibit thresholds close to the belief propagation algorithm. The class of two-bit decoders that we propose is a generalization of their idea, since we consider four possible values for the decoder messages.

Since the main focus of the paper is to establish sufficient conditions for correction of three errors, we do not optimize the decoders, but instead choose a specific decoder. Also, for the sake of simplicity we only consider universal decoders, i.e., decoders which do not depend on the channel parameter $\alpha$.

The rest of the paper is organized as follows. In Section II, we establish the notation and define a general class of two-bit decoders. For a specific two-bit decoder, the sufficient conditions for correction of three errors are derived in Section III. In Section IV, we derive thresholds for various decoders. Simulation results in Section V illustrate that, for a given code, lower FER can be achieved by a two-bit decoder compared to the FER achieved by Gallager B algorithm.

%% file: sec2.tex
\section{The class of two-bit decoders}
The Tanner graph of a code, whose parity-check matrix $\mathbf{H}$ has size $m\times n$, is a bipartite graph with a set of $n$ variable nodes and a set of $m$ check nodes. Each variable node corresponds to a column of the parity-check matrix, and each check node corresponds to a row. An edge connects a variable node to a check node if the corresponding element in the parity-check matrix is non-zero. A Tanner graph is said to be $\gamma$-left regular if all variable nodes have degree $\gamma$, $\rho$-right regular if all check nodes have degree $\rho$, and $(n,\gamma,\rho)$ regular if there are $n$ variable nodes, all variable nodes have degree $\gamma$ and all check nodes have degree $\rho$.

Gallager type algorithms for decoding over the BSC run iteratively. Let $\mathbf{r}$ be a binary $n$-tuple input to the decoder.
In the first half of each iteration, each variable node sends a message to its neighboring check nodes. The outgoing message along an edge depends on all the incoming messages except the one coming on that edge and possibly the received value. At the end of each iteration, a decision on the value of each bit is made in terms of all the messages going into the corresponding variable node.

Let $\omega_j(v,c)$ be the message that a variable node $v$ sends to its neighboring check node $c$ in the first half of the $j^{th}$ iteration. Analogously, $\varpi_j(c,v)$ denotes the message that a check node $c$ sends to its neighboring variable node $v$ in the second half of the $j^{th}$ iteration. Additionally, we define $\omega_j(v,:)$ as the set of all messages from a variable $v$ to all its neighboring checks at the beginning of the $j^{th}$ iteration. We define $\omega_j(v,:\backslash c)$ as the set of all messages that a variable node $v$ sends at the beginning of the $j^{th}$ iteration to all its neighboring checks except $c$. The sets $\varpi_j(c,:)$ and $\varpi_j(c,:\backslash v)$ are similarly defined.

\textit{Remark:} Since the message alphabet is finite, the message passing update rules can be described using a lookup table and hence only a finite number of two-bit decoders are possible. We assume two kinds of symmetry for the considered decoder. First, the Boolean function that represents any particular decoder must be symmetric in the sense that swapping all inputs must imply a swap of the output, i.e., the decoder performance does not depend on the sent codeword. Secondly, we consider only symmetric Boolean functions whose value depends only on the weight in the argument vector, not on positions of zeros and ones. Such symmetric Boolean functions are natural choice for regular codes. For irregular codes, asymmetric Boolean functions may lead to improved decoders, but this problem is out of the scope of this paper. In this paper, we focus on a class of two-bit decoders that can be described using simple algebraic rules and illustrate with an example how the lookup table can be constructed from the algebraic description.

%These symmetric rules can be seen as follows. The messages are of two kinds: strong and weak. One of the two bits of a message going into a variable node corresponds to the value of this variable node this message votes for, from a majority decoding point of view. The other bit determines the kind of the message. A strong message has a higher number of votes than a weak message. At the variable node, the votes of incoming messages, except the one being computed, are summed up. The value of the variable the outgoing message will carry is determined by the value getting the highest number of votes, while the strength of the outgoing message is determined by this number of votes.

Let the message alphabet be denoted by $M=\{-S,-W,W,S\}$ where $-S$ denotes a strong ``1'', $-W$ denotes a weak ``1'', $W$ denotes a weak ``0'', $S$ denotes a strong ``0'' and $S,W \in \mathbb{R}^+$. It should be noted that this representation can be mapped onto the alphabet $\{11,01,00,10\}$, but we use the symbols throughout for the sake of convenience. The received value $r_v \in \{0,1\}$ on the channel of a variable node $v$ is mapped to $R_v \in \{C,-C\}, C \in \mathbb{R}^+$, as follows: $1 \rightarrow -C$ and $0 \rightarrow C$. It can be seen that each message is associated with a value and strength (strength of a message is an indication of its reliability). 

Let $\mathcal{N}_1(u)$ denote the set of nodes connected to node $u$ by an edge.
Let the quantities $t_j(v,c)$ and $t_j(v)$, $j>1$ be defined as follows:
%\[t_j(v,c)=\sum_{u \in \mathcal{N}_1(v) \backslash c} \varpi_{j-1}(u,v)+R_v\]
%and
\begin{equation}\label{decisiont}
t_j(v,c)=\sum_{u \in \mathcal{N}_1(v) \backslash c} \varpi_{j-1}(u,v)+R_v\quad,\quad t_j(v)=\sum_{u \in \mathcal{N}_1(v)} \varpi_{j}(u,v)+R_v
\end{equation}
Additionally, let \[sign(\varpi_j(c,v))=\prod_{u \in \mathcal{N}_1(c) \backslash v}sign(\omega_j(u,c)),\]
where $sign(a)=1$, if $a\geq0$ and $sign(a)=-1$, if $a<0$.

The message passing update and decision rules can be expressed as follows. The absolute value is denoted by $|\cdot|$.
\[\omega_1(v,c)=W\cdot sign(R_v)\quad,\quad
\varpi_j(c,v)=
\left\{
\begin{array}{ll}
S\cdot sign(\varpi_j(c,v)) ,&\mbox{if } \forall u \in \mathcal{N}_1(c) \backslash v,\\
&|\omega_j(u,c)|=S\\ \\
W\cdot sign(\varpi_j(c,v)),&\mbox{otherwise}
\end{array}
\right.\]
For $j > 1$:
\[
\omega_j(v,c) =
\left\{
\begin{array}{ll}
W\cdot sign(t_j(v,c)),&\mbox{if } 0<|t_j(v,c)|<S\\
 & \\
S\cdot sign(t_j(v,c)),&\mbox{if } |t_j(v,c)|\geq S\\
 & \\
W\cdot sign(R_v),&\mbox{if } t_j(v,c)=0
\end{array}
\right. 
\]
\textit{Decision:} At the end of $j^{th}$ iteration, the estimate $r_v^j$ of a variable node $v$ is given by
\begin{eqnarray*}
r^j_v &=&
\left\{
\begin{array}{ll}
0,&\mbox{if } t_j(v) > 0\\
 & \\
1,&\mbox{if } t_j(v) < 0\\
 & \\
r_v ,&\mbox{if } t_j(v)=0
\end{array}
\right.
\end{eqnarray*}

The class of two-bit decoders described above can be interpreted as a voting scheme in the following way: every message has two components, namely the value (0 or 1) and the strength (weak or strong). The sign of the message determines the value, whereas the values of $W$ and $S$ denote the number of votes. The received value is associated with $C$ votes. To compute the outgoing message on the variable node side, the total number of votes corresponding to $0$ and $1$ are summed. The value of the outgoing message is the bit with more number of votes and the strength is determined by the number of votes. In the case of a tie, the outgoing message is set to the received value with a weak strength. 
Table \ref{example} gives an example of message update for a column-weight-four code, when $C=2$, $S=2$ and $W=1$. The message $\omega_j(v,c)$ goes out of variable node $v$, and is computed in terms of the three messages going into $v$ from the neighboring check nodes different of $c$. Table \ref{tabmp} shows the message passing update rules for $(C,S,W)=(2,2,1)$. Table \ref{tabdec} shows the decision rules for $(C,S,W)=(2,2,1)$.

\begin{table}
\begin{center}
\caption{Examples of message update for a column-weight-four code, when $C=2$, $S=2$ and $W=1$.}
\label{example}
\begin{tabular}{|c||c|c|c|c|}\hline
$\#$ incoming $-S$ messages & 2 & 1 & 0 & 1\\ \hline
$\#$ incoming $-W$ messages & 0 & 1 & 2 & 0\\ \hline
$\#$ incoming $W$ messages & 1 & 0 & 0 & 1\\ \hline
$\#$ incoming $S$ messages & 0 & 1 & 1 & 1\\ \hline
$R_v$ & $-C$ & $C$ & $C$ & $-C$\\ \hline
$\omega_j(v,c)$ & $-S$ & $W$ & $S$ & $-W$ \\ \hline
\end{tabular}
\end{center}
\end{table}

Different decoders in this class can be obtained by varying the values of $S,W$ and $C$. Hence, we denote a particular decoder by the triplet $(C,S,W)$. Since there are only a finite number of two-bit decoders, different choices for $C,S$ and $W$ might lead to the same decoder. Let $\mathcal{C}$ denote the class of above algebraically described decoders. Let us consider the set of all possible two-bit decoders which are symmetric in the senses that the performance is the same for any codeword, and whose Boolean functions do not depend on the location of $0$ and $1$ in the entries, but only on the weight. Let $\mathcal{S}$ denote such a set of rules. Then the following question arises: is this set $\mathcal{S}$ encompassed in $\mathcal{C}$? We do not intend to formally address this question, but rather give a discussion.

The answer is obviously no. However, among all these rules in $\mathcal{S}$, only a few are decoders, in the sense that if no error occurred, the sent codeword is output. Among these latter rules, only a few are further capable of correcting errors. We define the quality of a given rule by its threshold of convergence $\alpha^{\star}$, which is the maximum crossover probability of the BSC for which it is possible to achieve an arbitrary small error probability under iterative decoding, as the codeword length tends to infinity. Thresholds of two-bit decoders are further discussed in Section \ref{sec_asymp}. In order to verify that the rules in $\mathcal{C}$ allow to reach the best possible thresholds achievable with general two-bit decoders, we empirically checked that for any rule in $\mathcal{S}\backslash\mathcal{C}$ with better threshold than a rule in $\mathcal{C}$ with reasonable threshold, there exists a rule in $\mathcal{C}$ which has an at least as good threshold. Hence, we did an exhaustive scan of possible rules, for two $(d_v,d_c)$ regular code ensembles: $(3,4)$ and $(4,5)$, where $d_v$ and $d_c$ are the connection degrees of variable and check nodes, respectively. It is observed that for the $(4,5)$ regular code ensemble, no rule in $\mathcal{S}\backslash\mathcal{C}$ has better threshold than any rule in $\mathcal{C}$. For the $(3,4)$ regular code ensemble, only two rules in $\mathcal{S}\backslash\mathcal{C}$ have better threshold than any rule in $\mathcal{C}$, but with a very slight difference ($0.078$ for the former versus $0.075$ for the latter). Thus, it is reasonable to assume the class $\mathcal{C}$ of algebraically described decoders are representative of the best possible two-bit decoders.

In the next section, we focus on the two-bit decoder with $(C,S,W)=(2,2,1)$, and provide the conditions on the Tanner graph of the code to correct all patterns with up to three errors. As shown in Section \ref{sec_asymp}, this decoder has better thresholds than one-bit decoders for various code rates.

%% file: sec4.tex
\section{Guaranteed weight-three error correction}
In this section, we first find sufficient conditions on the Tanner graph of a code to ensure that the code can correct up to three errors in the codeword, when the decoding is performed with the two-bit decoder with $(C,S,W)=(2,2,1)$. As justified in the introduction, we consider only left-regular codes with column weight four. 

%\begin{figure*}
\begin{table}[!htb]
\caption{Decision rule: Number of messages $-S$, $-W$, $W$ and $S$ going into a variable, when this variable node is decoded as 0 (resp. 1) when the channel observation is 1 (resp. 0). The code has column weight four and the two-bit decoder has $(C,S,W)=(2,2,1)$.}
\label{tabdec}
\begin{center}
{\footnotesize
\begin{tabular}{|c|c|c|c|c|}\hline
 & $\#$ $-S$ & $\#$ $-W$ & $\#$ $W$ & $\#$ $S$\\
 & mess. & mess. & mess. & mess.\\ \hline
 & $0$ & $0$ & $0$ & $4$\\ 
 & $0$ & $0$ & $1$ & $3$\\ 
 & $0$ & $0$ & $2$ & $2$\\ 
Received value 1 & $0$ & $0$ & $3$ & $1$\\ 
Decoded as 0 & $0$ & $0$ & $4$ & $0$\\ 
 & $0$ & $1$ & $0$ & $3$\\ 
 & $0$ & $1$ & $1$ & $2$\\ 
 & $0$ & $1$ & $2$ & $1$\\ 
 & $1$ & $0$ & $0$ & $3$\\ 
 & $1$ & $0$ & $1$ & $2$\\ \hline
 & $0$ & $4$ & $0$ & $0$\\ 
 & $1$ & $2$ & $1$ & $0$\\ 
 & $1$ & $3$ & $0$ & $0$\\ 
Received value 0 & $2$ & $1$ & $0$ & $1$\\ 
Decoded as 1 & $2$ & $1$ & $1$ & $0$\\ 
 & $2$ & $2$ & $0$ & $0$\\ 
 & $3$ & $0$ & $0$ & $1$\\ 
 & $3$ & $0$ & $1$ & $0$\\ 
 & $3$ & $1$ & $0$ & $0$\\ 
 & $4$ & $0$ & $0$ & $0$\\ \hline
\end{tabular}}
\end{center}
\end{table}

\begin{table}[!htb]
\caption{Update rule: Number of messages $-S$, $-W$, $W$ and $S$ going into the variable node $v$ leading to different values of the message $\omega_j(v,c)$ going out of $v$, when the received value is $r_v$. The code has column weight four and the two-bit decoder has $(C,S,W)=(2,2,1)$.}
\label{tabmp}
\begin{center}
{\footnotesize
\begin{tabular}{|c|c|c|c|c|}\hline
 & $\#$ $-S$ & $\#$ $-W$ & $\#$ $W$ & $\#$ $S$\\
 & mess. & mess. & mess. & mess.\\ \hline
$r_v=1$ & $0$ & $0$ & $2$ & $1$\\ 
$\omega_j(v,c)=W$ & $0$ & $0$ & $3$ & $0$\\ 
 & $0$ & $1$ & $0$ & $2$\\ \hline
$r_v=1$ & $0$ & $0$ & $0$ & $3$\\ 
$\omega_j(v,c)=S$ & $0$ & $0$ & $1$ & $2$\\ \hline
$r_v=0$ & $2$ & $1$ & $0$ & $0$\\ 
$\omega_j(v,c)=-S$ & $3$ & $0$ & $0$ & $0$\\ \hline
 & $0$ & $3$ & $0$ & $0$\\ 
$r_v=0$ & $1$ & $2$ & $0$ & $0$\\ 
$\omega_j(v,c)=-W$ & $2$ & $0$ & $1$ & $0$\\ \hline
 & $0$ & $2$ & $0$ & $1$\\ 
 & $0$ & $2$ & $1$ & $0$\\ 
 & $0$ & $3$ & $0$ & $0$\\ 
 & $1$ & $0$ & $2$ & $0$\\ 
 & $1$ & $1$ & $0$ & $1$\\ 
$r_v=1$ & $1$ & $1$ & $1$ & $0$\\ 
$\omega_j(v,c)=-S$ & $1$ & $2$ & $0$ & $0$\\ 
 & $2$ & $0$ & $0$ & $1$\\ 
 & $2$ & $0$ & $1$ & $0$\\ 
 & $2$ & $1$ & $0$ & $0$\\ 
 & $3$ & $0$ & $0$ & $0$\\ \hline
 & $0$ & $1$ & $1$ & $1$\\ 
$r_v=1$ & $0$ & $1$ & $2$ & $0$\\ 
$\omega_j(v,c)=-W$ & $1$ & $0$ & $0$ & $2$\\ 
 & $1$ & $0$ & $1$ & $1$\\ \hline
 & $0$ & $2$ & $1$ & $0$\\ 
$r_v=0$ & $1$ & $1$ & $0$ & $1$\\ 
$\omega_j(v,c)=W$ & $1$ & $1$ & $1$ & $0$\\ 
 & $2$ & $0$ & $0$ & $1$\\ \hline
 & $0$ & $0$ & $0$ & $3$\\ 
 & $0$ & $0$ & $1$ & $2$\\ 
 & $0$ & $0$ & $2$ & $1$\\ 
 & $0$ & $0$ & $3$ & $0$\\ 
$r_v=0$ & $0$ & $1$ & $0$ & $2$\\ 
$\omega_j(v,c)=S$ & $0$ & $1$ & $1$ & $1$\\ 
 & $0$ & $1$ & $2$ & $0$\\ 
 & $0$ & $2$ & $0$ & $1$\\ 
 & $1$ & $0$ & $0$ & $2$\\ 
 & $1$ & $0$ & $1$ & $1$\\ 
 & $1$ & $0$ & $2$ & $0$\\ \hline
\end{tabular}}
\end{center}
\end{table}
%\end{figure*}

Since the code is linear and the channel and the decoder are symmetric, we can assume, without loss of generality, that the all-zero codeword is transmitted over the BSC. We make this assumption throughout the paper. Hence, the variable nodes flipped by the channel are received as ``1''. 

The problem of guaranteed error correction capability assumes significance in the error floor region. Roughly speaking, error floor is the abrupt degradation in the FER performance in the high SNR regime. The error floor phenomenon has been attributed to the presence of a few harmful configurations in the Tanner graph of the code, variously known as stopping sets (for the BEC), near codewords \cite{03MP}, trapping sets (for iterative decoding on the BSC and the AWGN) and pseudo-codewords (for linear programming decoding) \cite{05FWK}. While girth optimized codes have been known to perform well in general, the code length and the degree distribution place a fundamental limit on the best achievable girth. Hence, additional constraints on the Tanner graph are required to ensure better error floor performance. 

The guaranteed error correction capability of column-weight-three LDPC codes under the Gallager A algorithm is now completely understood (see \cite{08CV,08CNVM2} for details). For column-weight-four LDPC codes under the Gallager B algorithm, sufficient conditions to guarantee the correction of all error patterns with up to three errors have been derived by Chilappagari \textit{et al.}\cite{08CKVM}. The conditions derived in \cite{08CKVM} impose constraints on the least number of neighboring check nodes for a given set of variable nodes. The conditions that we derive are similar, but impose fewer constraints on the Tanner graph, thereby resulting in codes with higher rates for the same length. A short discussion on this issue is provided at the end of the section.

Let us first give some additional definition and notation.
\begin{defi}
The neighborhood of depth one of a node $u$ is denoted by $\mathcal{N}_1(u)$ and is composed of all the nodes such that there exists an edge between these nodes and $u$. Similarly, $\mathcal{N}_d(u)$ denotes the neighborhood of depth $d$ of node $u$ and is composed of all the nodes such that there exists a path of length $d$ between these nodes and $u$.
\end{defi}

Let $E$ be a set of nodes, say $E=\cup_i u_i$, then the depth $d$ neighborhood of $E$ is $\mathcal{N}_d(E)=\cup_i \mathcal{N}_d(u_i)$. %Let $V^1=\{v_1^1,v_2^1,v_3^1\}$ and $C^1=\mathcal{N}_1(V^1)$. For more easily readable notations, 

Now we state the main theorem.

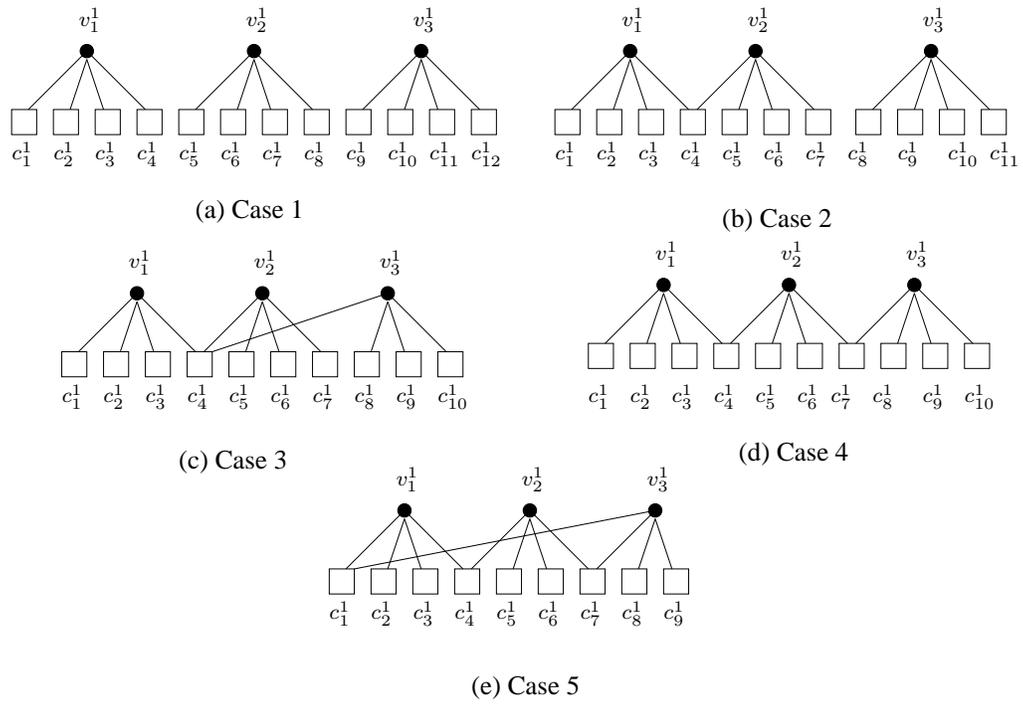
\begin{figure}[!hb]
\begin{center} \input{errors.pstex_t} \end{center}
\caption{All possible subgraphs subtended by three erroneous variable nodes.}
\label{errors}
\end{figure}

\begin{theo}\label{th}[Irregular expansion theorem]
Let $\mathcal{G}$ be the Tanner graph of a column-weight-four LDPC code with no 4-cycles, satisfying the following expansion conditions: each variable subset of size 4 has at least 11 neighbors, each one of size 5 at least 12 neighbors, each one of size 6 at least 14 neighbors, each one of size 8 at least 16 neighbors and each one of size 9 at least 18 neighbors. The two-bit decoder, with $C=2$, $S=2$ and $W=1$, can correct up to three errors in the codeword within three iterations, if and only if the above conditions are satisfied.
\end{theo}
For ease in notation, each expansion condition will be denoted by ``4$\rightarrow$11 expansion condition'', ``5$\rightarrow$12 expansion condition'' and so on.\\
\textit{Proof of sufficiency}:\\
\textit{Remark}: The proof can be followed more easily by looking at Tables \ref{tabdec} and \ref{tabmp}. Let $V^1=\{v_1^1,v_2^1,v_3^1\}$ and $C^1=\mathcal{N}_1(V^1)$. For more easily readable notation, let $\mathcal{N}_2(V^1)\backslash V^1$ be denoted by $V^2$ and $\mathcal{N}_1(V^2)\backslash C^1$ by $C^2$. Also, we say that a variable node is of type $T_p^q$ when it has $p$ connections to $C^1$ and $q$ connection to $C^2$. The union of order $d$ neighborhoods of all the $T_p^q$ variable nodes is denoted by $\mathcal{N}_d(T_p^q)$.\\
We consider all the subgraphs induced by three erroneous variable nodes in a graph and prove that, in each case, the errors are corrected. The possible subgraphs are shown in Figure \ref{errors}. As shown, five cases arise. In the reminder, we assume that the all-zero codeword has been sent. We provide the proof for Case 4 and relegate the proofs for necessity and other cases to the Appendix.

%\begin{figure}[!htb]
%\begin{center}
%\input{fig1.pstex_t}
%\end{center}
%\caption{Errors configuration for Case 2.}
%\label{cas2}
%\end{figure}

\textbf{Case 4}: 
%\begin{figure}
%\begin{center}
%\input{fig2.pstex_t}
%\end{center}
%\caption{Errors configuration for Case 4.}
%\label{cas4}
%\end{figure}
Consider the error configuration shown in Figure \ref{errors}(d). 
In the second half of the first iteration we have:
{\small
\begin{eqnarray}
\varpi_1(c,:\backslash V^1)&=&-W\quad,\quad c\in C^1\backslash \{c_4^1,c_7^1\}\nonumber\\
\varpi_1(c,v)&=&-W\quad,\quad v\in V^1,c\in \{c_4^1,c_7^1\}\nonumber\\
\varpi_1(c,v)&=&W\quad,\quad \mbox{otherwise} \nonumber
\end{eqnarray}}
Let us analyze the second iteration. For any $v\in V\backslash V^1$ and $c\in C^1$, $\omega_2(v,c)$ can never be $-S$ because no $-S$ messages propagate in the first iteration. So, for any $v\in V\backslash V^1$ and $c\in C^1$, $\omega_2(v,c)=-W$ if and only if $\varpi_1(:\backslash c,v)=-W$, which implies that $v$ must have four connections to $C^1$. This is not possible as it would cause a 4-cycle. Hence:
{\small
\begin{eqnarray}
\omega_2(v_2^1,c)&=&-S\quad,\quad c \in \{c_5^1,c_6^1\}\nonumber\\
\omega_2(v_2^1,c_4^1)&=&-W\nonumber\\
\omega_2(v_2^1,c_7^1)&=&-W\nonumber\\
\omega_2(v_1^1,:\backslash c_4^1)&=&-W\nonumber\\
\omega_2(v_3^1,:\backslash c_7^1)&=&-W\nonumber\\
\omega_2(v,c)&=&-W\quad v\in \mathcal{N}_0(T_3^1),\quad c\in C^2\cap \mathcal{N}_1(T_3^1)\nonumber\\
\omega_2(v_1^1,c_4^1)&=&W\nonumber\\
\omega_2(v_3^1,c_7^1)&=&W\nonumber\\
\omega_2(v,c)&=&W\quad v\in \mathcal{N}_0(T_2^2),\quad c\in C^2\cap\mathcal{N}_1(T_2^2)\nonumber\\
\omega_2(v,c)&=&W\quad v\in \mathcal{N}_0(T_3^1),\quad c\in C^1\cap\mathcal{N}_1(T_3^1)\nonumber\\
\omega_2(v,c)&=&S\quad,\quad \mbox{otherwise}\nonumber
%%%%%%%%%
%\varpi_2(c_4^1,:\backslash v_2^1)&=&-W\nonumber\\
%\varpi_2(c_7^1,:\backslash v_2^1)&=&-W\nonumber\\
%\varpi_2(c,:\backslash V^1)&=&-W\quad,\quad c\in C^1\nonumber\\
%\varpi_2(c,v)&=&W\quad,\quad \{c,v\}\in C^1\times V^1\backslash\{\{c_4^1,v_1^1\},\{c_7^1,v_3^1\}\}\nonumber
%%%%%%
\end{eqnarray}}
In the first half of the third iteration, we have
{\small
\begin{eqnarray}
\omega_3(v_2^1,:)&=&W\nonumber\\
\omega_3(v_1^1,:\backslash c_4^1)&=&-W\quad,\quad \omega_3(v_1^1,c_4^1)=W\nonumber\\
\omega_3(v_3^1,:\backslash c_7^1)&=&-W\quad,\quad \omega_3(v_3^1,c_7^1)=W\nonumber
\end{eqnarray}}
\begin{lemma}
All variables in $V^1$ are corrected at the end of the third iteration because, for any $v\in V^1$, $\varpi_3(:,v)=W$ or $S$.
\end{lemma}
\textit{Proof}: The proof is by contradiction. Let us assume that there exists a variable in $V\backslash V^1$, say $v$, such that there exists $c\in C^1$ and $\omega_3(v,c)=-W$ or $\omega_3(v,c)=-S$. Since it is impossible that two $-S$ messages go into $v$, as it would cause a 4-cycle, $\omega_3(v, c) = -W$ or $\omega_3(v, c) = -S$ implies that $v$ receives from its neighbors different of $c$, at the end of the second iteration, three $-W$ messages, or one $-S$ and two $-W$ (see Table \ref{tabmp}).

\begin{itemize}
\item If $v$ receives three $-W$: As proved previously, $v$ cannot have four neighbors in $C^1$. Hence, $v$ must be connected to $c_1^2\in C^2$ such that $\varpi_2(c_1^2,v)=-W$. With the above described values of the messages in the second half of the second iteration, we see that $c_1^2$ must be connected to a $T_3^1$ variable in $V^2$, say $x_1^2$. Let notice that there cannot be more than two $T_3^1$ variables in $V^2$, otherwise six variables would be connected to only thirteen checks. We are interested in $v$ which has at least one connection to $C^1$. $v$ has at most three connections to $C^1$. Three cases arise:
	\begin{itemize}
	\item If $v$ has three connections to $C^1$, then $v$ must have one neighboring check in $C^2$, say $c_1^2$, which has at least one neighboring variable, say $v'$, in $\mathcal{N}_0(T_3^1)$ different of $v$. Then the set $\{v_1^1,v_2^1,v_3^1,v,v'\}$ has only eleven neighbors, therefore contradicting the $5\rightarrow 12$ expansion condition.
	\item If $v$ has at two connections to $C^1$, then $v$ has two neighboring checks in $C^2$, say $c_1^2$ and $c_2^2$, which must have each at least one neighboring variable, say $v'$ and $v''$, in $\mathcal{N}_0(T_3^1)$ different of $v$. Then the set $\{v_1^1,v_2^1,v_3^1,v,v',v''\}$ has only twelve neighbors, therefore contradicting the $6\rightarrow 14$ expansion condition.
	\item If $v$ has at only one connection to $C^1$, then $v$ must have three neighboring checks in $C^2$, each of them connected to a $T_3^1$ variable. This has been previously proved to be impossible.
	\end{itemize}
\item If $v$ receives two $-W$ messages and one $-S$ message:
	\begin{itemize}
	\item If $v$ has at three connections to $C^1$, then we end up in the same situation as in the first item, where the $5\rightarrow 12$ expansion condition is not satisfied.
	\item If $v$ has at two connections to $C^1$ (one to $\{c_5^1,c_6^1\}$ to receive a $-S$ message, the other one to propagate a $-W$ or $-S$ message to $C^1$), then we end up in the same situation as in the first item, where the $6\rightarrow 14$ expansion condition is not satisfied.
	\end{itemize}
\end{itemize}

Hence, $v_1^1,v_2^1$ and $v_3^1$ are corrected at the end of the third iteration.
\begin{flushright}$\blacksquare$\end{flushright}
\begin{lemma}\label{lem}
No variable in $V\backslash V^1$ can propagate $-W$ at the beginning of the third iteration, except variables of type $T_3^1$, and $T_2^2$ variables which have a common check node in $C^2$ with a $T_3^1$ variable.
\end{lemma}
\textit{Proof}: 
\begin{itemize}
\item Consider a variable $v$ which has two connections to $C^1$. For this variable $v$ to propagate $-W$ at the beginning of the third iteration, it is necessary to receive a $-S$ or $-W$ message from one of its two check nodes in $C^2$, which is the case only if it shares a check node in $C^2$ with a $T_3^1$ variable.
\item Consider a variable $v$ which has exactly one connection to $C^1$. For this variable $v$ to propagate $-W$ at the beginning of the third iteration, it is necessary to receive a $-S$ or $-W$ message from two of its three check nodes in $C^2$, say $c_1^2$ and $c_2^2$, which is the case only if $c_1^2$ and $c_2^2$ are both shared by $T_3^1$ variables, say $v_1^2$ and $v_2^2$. Then the set $\{v_1^1,v_2^1,v_3^1,v_1^2,v_2^2,v\}$ is connected to only 12 checks, therefore contradicting the $6\rightarrow 14$ expansion condition.
\item Consider a variable $v$ which has no connection to $C^1$. For this variable $v$ to propagate $-W$ at the beginning of the third iteration, it is necessary to receive a $-S$ or $-W$ message from three of its four check nodes in $C^2$. This implies the existence of three $T_3^1$ variables, which has already been proved to be impossible.
\end{itemize}
\begin{flushright}$\blacksquare$\end{flushright}
\begin{lemma}
Any variable in $V\backslash V^1$ is correctly decoded at the end of the third iteration. 
\end{lemma}
\textit{Remark}: That is to say that any variable in $V\backslash V^1$ is decoded to its received value since it is not received in error by hypothesis.
\textit{Proof}: According to Table \ref{tabmp}, no message $-S$ propagates in the third iteration since all variables in $V^1$ receive at least three $W$ messages at the end of the second iteration, and variables in $V\backslash V^1$ cannot receive more than one $-S$ message. In that case, to be decoded as a one, a bit whose received value is zero has to receive only $-W$ messages according to the decision rule (see Table \ref{tabdec}). That is for any $v \in V\backslash V^1$, $v$ is wrongly decoded if and only if $\varpi_3(:, v) = -W$. 
Let $E$ denote the set of $T_2^2$ variables which share a check in $C^2$ with a $T_3^1$ variable.

Firstly, let consider a variable in $E$, say $v$, and let us call $v'$ the $T_3^1$ variable with which $v$ shares a check node in $C^2$. There cannot exist in the graph, at the same time, $v$ and a $T_3^1$ variable, say $v''$, different of $v'$. If such variables would exist, $v$, $v'$, $v''$ and the variables in $V^1$ would be connected to only 13 check nodes, therefore contradicting the $6\rightarrow 14$ expansion condition.
Secondly, no $v \in V\backslash V^1$ can have more than two neighboring checks in $\{c_1^1,c_2^1,c_3^1,c_8^1,c_9^1,c_{10}^1\}$, otherwise it would introduce a 4-cycle.
Hence, only de following cases are possible for a variable $v$ not in $V^1$ to receive four wrong messages:
\begin{itemize}
\item If $v$ has no connection to $E$, two cases arise:
\begin{itemize}
\item If $v$ has two connections in $\{c_1^1,c_2^1,c_3^1,c_8^1,c_9^1,c_{10}^1\}$ and two connections to $\mathcal{N}_1(T_3^1)\cap C^2$. Then $v$, the variables in $V^1$ and the two $T_3^1$ variables are connected to only 12 check nodes, therefore contradicting the $6\rightarrow 14$ expansion condition.
\item If $v$ has at most one connection to $\{c_1^1,c_2^1,c_3^1,c_8^1,c_9^1,c_{10}^1\}$, it must have at least three connections to $\mathcal{N}_1(T_3^1)\cap C^2$. However, there cannot exist three $T_3^1$ variables as it would imply that the set made of these three variables and $V^1$ would be connected to only 13 check nodes, therefore contradicting the $6\rightarrow 14$ expansion condition.
\end{itemize}
\item If $v$ has no connection to $\mathcal{N}_1(T_3^1)\cap C^2$, two cases arise:
\begin{itemize}
\item If $v$ has two connections in $\{c_1^1,c_2^1,c_3^1,c_8^1,c_9^1,c_{10}^1\}$ and two connections to $E$. Let consider one of the two variables in $T_2^2$, say $v'$, the $T_3^1$ variable with which $v'$ shares a check node in $C^2$, $v$ and the variables in $V^1$. Then this set of variables is connected only to 13 check nodes, therefore contradicting the $6\rightarrow 14$ expansion condition.
\item If $v$ has at most one connection to $\{c_1^1,c_2^1,c_3^1,c_8^1,c_9^1,c_{10}^1\}$, it must have at least three connections to $E$. This implies the existence of three distinct $T_3^1$ variables, which is impossible, as above mentioned.
\end{itemize}
\end{itemize}
\begin{flushright}$\blacksquare$\end{flushright}
Thus, the decoder converges to the valid codeword at the end of the third iteration.\\
Note that similar conditions for a column-weight-four LDPC code of girth six to correct any weight-three error pattern within four iterations, when it is decoded with Gallager B algorithm, has been found by Chilappagari \textit{et al.} \cite{08CKVM}. The conditions are that each variable subset of size 4 has at least 11 neighbors, each one of size 5 at least 12 neighbors, each one of size 6 at least 14 neighbors, each one of size 7 at least 16 neighbors and each one of size 8 at least 18 neighbors. These conditions are stronger than the ones of Theorem \ref{th} in two aspects, on which we wish to have a short discussion.\\
On one hand, provided that the respective graph conditions are fulfilled, the number of required iterations to correct three errors is lower for the $(2,2,1)$ two-bit decoder than for the Gallager B decoder. However, since messages are quantified over two bits for the former and over one bit for the latter, a lower number of iterations does not necessarily mean a lower decoding complexity. We do not provide here further analysis for comparison of decoding complexity between both kinds of decoding, as it would highly depend on hardware choices.

On the other hand, the higher the rate of the code, the more difficult for the Tanner graph of the code to satisfy the expansion conditions, since the variable nodes tend to be less and less connected when the code rate increases. Hence, it is likely that weaker expansion conditions, obtained for the two-bit decoder, make possible the construction of higher rate codes, with weight-three error correction capability, than expansion conditions required by the one-bit Gallager B decoder do. However, determining analytically the highest achievable rate for a given set of expansion conditions is a problem which may be very hard to solve, and which is out of the scope of this paper.

%ensure that weight-three error correction capability is achievable for higher rate codes when the decoding is performed with the two-bit decoder, than with the one-bit Gallager B decoder.

%\subsection{Code criterion construction for correction of weight-three errors}

%% file: errors.pstex_t
\begin{picture}(0,0)%
\includegraphics{errors.pstex}%
\end{picture}%
\setlength{\unitlength}{2763sp}%
\begingroup\makeatletter\ifx\SetFigFont\undefined%
\gdef\SetFigFont#1#2#3#4#5{%
  \reset@font\fontsize{#1}{#2pt}%
  \fontfamily{#3}\fontseries{#4}\fontshape{#5}%
  \selectfont}%
\fi\endgroup%
\begin{picture}(8952,6249)(286,-6214)
\put(6001,-4261){\makebox(0,0)[lb]{\smash{{\SetFigFont{8}{9.6}{\rmdefault}{\mddefault}{\updefault}{\color[rgb]{0,0,0}$v_3^1$}%
}}}}
\put(301,-1336){\makebox(0,0)[lb]{\smash{{\SetFigFont{8}{9.6}{\rmdefault}{\mddefault}{\updefault}{\color[rgb]{0,0,0}$c_1^1$}%
}}}}
\put(676,-1336){\makebox(0,0)[lb]{\smash{{\SetFigFont{8}{9.6}{\rmdefault}{\mddefault}{\updefault}{\color[rgb]{0,0,0}$c_2^1$}%
}}}}
\put(1051,-1336){\makebox(0,0)[lb]{\smash{{\SetFigFont{8}{9.6}{\rmdefault}{\mddefault}{\updefault}{\color[rgb]{0,0,0}$c_3^1$}%
}}}}
\put(1426,-1336){\makebox(0,0)[lb]{\smash{{\SetFigFont{8}{9.6}{\rmdefault}{\mddefault}{\updefault}{\color[rgb]{0,0,0}$c_4^1$}%
}}}}
\put(1801,-1336){\makebox(0,0)[lb]{\smash{{\SetFigFont{8}{9.6}{\rmdefault}{\mddefault}{\updefault}{\color[rgb]{0,0,0}$c_5^1$}%
}}}}
\put(2176,-1336){\makebox(0,0)[lb]{\smash{{\SetFigFont{8}{9.6}{\rmdefault}{\mddefault}{\updefault}{\color[rgb]{0,0,0}$c_6^1$}%
}}}}
\put(2551,-1336){\makebox(0,0)[lb]{\smash{{\SetFigFont{8}{9.6}{\rmdefault}{\mddefault}{\updefault}{\color[rgb]{0,0,0}$c_7^1$}%
}}}}
\put(2926,-1336){\makebox(0,0)[lb]{\smash{{\SetFigFont{8}{9.6}{\rmdefault}{\mddefault}{\updefault}{\color[rgb]{0,0,0}$c_8^1$}%
}}}}
\put(3301,-1336){\makebox(0,0)[lb]{\smash{{\SetFigFont{8}{9.6}{\rmdefault}{\mddefault}{\updefault}{\color[rgb]{0,0,0}$c_9^1$}%
}}}}
\put(3676,-1336){\makebox(0,0)[lb]{\smash{{\SetFigFont{8}{9.6}{\rmdefault}{\mddefault}{\updefault}{\color[rgb]{0,0,0}$c_{10}^1$}%
}}}}
\put(4051,-1336){\makebox(0,0)[lb]{\smash{{\SetFigFont{8}{9.6}{\rmdefault}{\mddefault}{\updefault}{\color[rgb]{0,0,0}$c_{11}^1$}%
}}}}
\put(4426,-1336){\makebox(0,0)[lb]{\smash{{\SetFigFont{8}{9.6}{\rmdefault}{\mddefault}{\updefault}{\color[rgb]{0,0,0}$c_{12}^1$}%
}}}}
\put(5176,-1336){\makebox(0,0)[lb]{\smash{{\SetFigFont{8}{9.6}{\rmdefault}{\mddefault}{\updefault}{\color[rgb]{0,0,0}$c_1^1$}%
}}}}
\put(5551,-1336){\makebox(0,0)[lb]{\smash{{\SetFigFont{8}{9.6}{\rmdefault}{\mddefault}{\updefault}{\color[rgb]{0,0,0}$c_2^1$}%
}}}}
\put(5926,-1336){\makebox(0,0)[lb]{\smash{{\SetFigFont{8}{9.6}{\rmdefault}{\mddefault}{\updefault}{\color[rgb]{0,0,0}$c_3^1$}%
}}}}
\put(6301,-1336){\makebox(0,0)[lb]{\smash{{\SetFigFont{8}{9.6}{\rmdefault}{\mddefault}{\updefault}{\color[rgb]{0,0,0}$c_4^1$}%
}}}}
\put(6676,-1336){\makebox(0,0)[lb]{\smash{{\SetFigFont{8}{9.6}{\rmdefault}{\mddefault}{\updefault}{\color[rgb]{0,0,0}$c_5^1$}%
}}}}
\put(7051,-1336){\makebox(0,0)[lb]{\smash{{\SetFigFont{8}{9.6}{\rmdefault}{\mddefault}{\updefault}{\color[rgb]{0,0,0}$c_6^1$}%
}}}}
\put(8701,-1336){\makebox(0,0)[lb]{\smash{{\SetFigFont{8}{9.6}{\rmdefault}{\mddefault}{\updefault}{\color[rgb]{0,0,0}$c_{10}^1$}%
}}}}
\put(751,-3511){\makebox(0,0)[lb]{\smash{{\SetFigFont{8}{9.6}{\rmdefault}{\mddefault}{\updefault}{\color[rgb]{0,0,0}$c_1^1$}%
}}}}
\put(1126,-3511){\makebox(0,0)[lb]{\smash{{\SetFigFont{8}{9.6}{\rmdefault}{\mddefault}{\updefault}{\color[rgb]{0,0,0}$c_2^1$}%
}}}}
\put(1501,-3511){\makebox(0,0)[lb]{\smash{{\SetFigFont{8}{9.6}{\rmdefault}{\mddefault}{\updefault}{\color[rgb]{0,0,0}$c_3^1$}%
}}}}
\put(1876,-3511){\makebox(0,0)[lb]{\smash{{\SetFigFont{8}{9.6}{\rmdefault}{\mddefault}{\updefault}{\color[rgb]{0,0,0}$c_4^1$}%
}}}}
\put(2251,-3511){\makebox(0,0)[lb]{\smash{{\SetFigFont{8}{9.6}{\rmdefault}{\mddefault}{\updefault}{\color[rgb]{0,0,0}$c_5^1$}%
}}}}
\put(2626,-3511){\makebox(0,0)[lb]{\smash{{\SetFigFont{8}{9.6}{\rmdefault}{\mddefault}{\updefault}{\color[rgb]{0,0,0}$c_6^1$}%
}}}}
\put(3001,-3511){\makebox(0,0)[lb]{\smash{{\SetFigFont{8}{9.6}{\rmdefault}{\mddefault}{\updefault}{\color[rgb]{0,0,0}$c_7^1$}%
}}}}
\put(3376,-3511){\makebox(0,0)[lb]{\smash{{\SetFigFont{8}{9.6}{\rmdefault}{\mddefault}{\updefault}{\color[rgb]{0,0,0}$c_8^1$}%
}}}}
\put(3751,-3511){\makebox(0,0)[lb]{\smash{{\SetFigFont{8}{9.6}{\rmdefault}{\mddefault}{\updefault}{\color[rgb]{0,0,0}$c_9^1$}%
}}}}
\put(4126,-3511){\makebox(0,0)[lb]{\smash{{\SetFigFont{8}{9.6}{\rmdefault}{\mddefault}{\updefault}{\color[rgb]{0,0,0}$c_{10}^1$}%
}}}}
\put(5476,-3511){\makebox(0,0)[lb]{\smash{{\SetFigFont{8}{9.6}{\rmdefault}{\mddefault}{\updefault}{\color[rgb]{0,0,0}$c_1^1$}%
}}}}
\put(5851,-3511){\makebox(0,0)[lb]{\smash{{\SetFigFont{8}{9.6}{\rmdefault}{\mddefault}{\updefault}{\color[rgb]{0,0,0}$c_2^1$}%
}}}}
\put(6226,-3511){\makebox(0,0)[lb]{\smash{{\SetFigFont{8}{9.6}{\rmdefault}{\mddefault}{\updefault}{\color[rgb]{0,0,0}$c_3^1$}%
}}}}
\put(6601,-3511){\makebox(0,0)[lb]{\smash{{\SetFigFont{8}{9.6}{\rmdefault}{\mddefault}{\updefault}{\color[rgb]{0,0,0}$c_4^1$}%
}}}}
\put(6976,-3511){\makebox(0,0)[lb]{\smash{{\SetFigFont{8}{9.6}{\rmdefault}{\mddefault}{\updefault}{\color[rgb]{0,0,0}$c_5^1$}%
}}}}
\put(7351,-3511){\makebox(0,0)[lb]{\smash{{\SetFigFont{8}{9.6}{\rmdefault}{\mddefault}{\updefault}{\color[rgb]{0,0,0}$c_6^1$}%
}}}}
\put(3151,-5461){\makebox(0,0)[lb]{\smash{{\SetFigFont{8}{9.6}{\rmdefault}{\mddefault}{\updefault}{\color[rgb]{0,0,0}$c_1^1$}%
}}}}
\put(3526,-5461){\makebox(0,0)[lb]{\smash{{\SetFigFont{8}{9.6}{\rmdefault}{\mddefault}{\updefault}{\color[rgb]{0,0,0}$c_2^1$}%
}}}}
\put(3901,-5461){\makebox(0,0)[lb]{\smash{{\SetFigFont{8}{9.6}{\rmdefault}{\mddefault}{\updefault}{\color[rgb]{0,0,0}$c_3^1$}%
}}}}
\put(4276,-5461){\makebox(0,0)[lb]{\smash{{\SetFigFont{8}{9.6}{\rmdefault}{\mddefault}{\updefault}{\color[rgb]{0,0,0}$c_4^1$}%
}}}}
\put(4651,-5461){\makebox(0,0)[lb]{\smash{{\SetFigFont{8}{9.6}{\rmdefault}{\mddefault}{\updefault}{\color[rgb]{0,0,0}$c_5^1$}%
}}}}
\put(5026,-5461){\makebox(0,0)[lb]{\smash{{\SetFigFont{8}{9.6}{\rmdefault}{\mddefault}{\updefault}{\color[rgb]{0,0,0}$c_6^1$}%
}}}}
\put(5401,-5461){\makebox(0,0)[lb]{\smash{{\SetFigFont{8}{9.6}{\rmdefault}{\mddefault}{\updefault}{\color[rgb]{0,0,0}$c_7^1$}%
}}}}
\put(5776,-5461){\makebox(0,0)[lb]{\smash{{\SetFigFont{8}{9.6}{\rmdefault}{\mddefault}{\updefault}{\color[rgb]{0,0,0}$c_8^1$}%
}}}}
\put(6151,-5461){\makebox(0,0)[lb]{\smash{{\SetFigFont{8}{9.6}{\rmdefault}{\mddefault}{\updefault}{\color[rgb]{0,0,0}$c_9^1$}%
}}}}
\put(8026,-3511){\makebox(0,0)[lb]{\smash{{\SetFigFont{8}{9.6}{\rmdefault}{\mddefault}{\updefault}{\color[rgb]{0,0,0}$c_8^1$}%
}}}}
\put(7651,-3511){\makebox(0,0)[lb]{\smash{{\SetFigFont{8}{9.6}{\rmdefault}{\mddefault}{\updefault}{\color[rgb]{0,0,0}$c_7^1$}%
}}}}
\put(8476,-3511){\makebox(0,0)[lb]{\smash{{\SetFigFont{8}{9.6}{\rmdefault}{\mddefault}{\updefault}{\color[rgb]{0,0,0}$c_9^1$}%
}}}}
\put(7426,-1336){\makebox(0,0)[lb]{\smash{{\SetFigFont{8}{9.6}{\rmdefault}{\mddefault}{\updefault}{\color[rgb]{0,0,0}$c_7^1$}%
}}}}
\put(7801,-1336){\makebox(0,0)[lb]{\smash{{\SetFigFont{8}{9.6}{\rmdefault}{\mddefault}{\updefault}{\color[rgb]{0,0,0}$c_8^1$}%
}}}}
\put(8251,-1336){\makebox(0,0)[lb]{\smash{{\SetFigFont{8}{9.6}{\rmdefault}{\mddefault}{\updefault}{\color[rgb]{0,0,0}$c_9^1$}%
}}}}
\put(9076,-1336){\makebox(0,0)[lb]{\smash{{\SetFigFont{8}{9.6}{\rmdefault}{\mddefault}{\updefault}{\color[rgb]{0,0,0}$c_{11}^1$}%
}}}}
\put(8851,-3511){\makebox(0,0)[lb]{\smash{{\SetFigFont{8}{9.6}{\rmdefault}{\mddefault}{\updefault}{\color[rgb]{0,0,0}$c_{10}^1$}%
}}}}
\put(901,-136){\makebox(0,0)[lb]{\smash{{\SetFigFont{8}{9.6}{\rmdefault}{\mddefault}{\updefault}{\color[rgb]{0,0,0}$v_1^1$}%
}}}}
\put(2401,-136){\makebox(0,0)[lb]{\smash{{\SetFigFont{8}{9.6}{\rmdefault}{\mddefault}{\updefault}{\color[rgb]{0,0,0}$v_2^1$}%
}}}}
\put(3901,-136){\makebox(0,0)[lb]{\smash{{\SetFigFont{8}{9.6}{\rmdefault}{\mddefault}{\updefault}{\color[rgb]{0,0,0}$v_3^1$}%
}}}}
\put(5776,-136){\makebox(0,0)[lb]{\smash{{\SetFigFont{8}{9.6}{\rmdefault}{\mddefault}{\updefault}{\color[rgb]{0,0,0}$v_1^1$}%
}}}}
\put(6901,-136){\makebox(0,0)[lb]{\smash{{\SetFigFont{8}{9.6}{\rmdefault}{\mddefault}{\updefault}{\color[rgb]{0,0,0}$v_2^1$}%
}}}}
\put(8476,-136){\makebox(0,0)[lb]{\smash{{\SetFigFont{8}{9.6}{\rmdefault}{\mddefault}{\updefault}{\color[rgb]{0,0,0}$v_3^1$}%
}}}}
\put(8326,-2236){\makebox(0,0)[lb]{\smash{{\SetFigFont{8}{9.6}{\rmdefault}{\mddefault}{\updefault}{\color[rgb]{0,0,0}$v_3^1$}%
}}}}
\put(7201,-2236){\makebox(0,0)[lb]{\smash{{\SetFigFont{8}{9.6}{\rmdefault}{\mddefault}{\updefault}{\color[rgb]{0,0,0}$v_2^1$}%
}}}}
\put(6076,-2236){\makebox(0,0)[lb]{\smash{{\SetFigFont{8}{9.6}{\rmdefault}{\mddefault}{\updefault}{\color[rgb]{0,0,0}$v_1^1$}%
}}}}
\put(3601,-2311){\makebox(0,0)[lb]{\smash{{\SetFigFont{8}{9.6}{\rmdefault}{\mddefault}{\updefault}{\color[rgb]{0,0,0}$v_3^1$}%
}}}}
\put(1351,-2311){\makebox(0,0)[lb]{\smash{{\SetFigFont{8}{9.6}{\rmdefault}{\mddefault}{\updefault}{\color[rgb]{0,0,0}$v_1^1$}%
}}}}
\put(2476,-2311){\makebox(0,0)[lb]{\smash{{\SetFigFont{8}{9.6}{\rmdefault}{\mddefault}{\updefault}{\color[rgb]{0,0,0}$v_2^1$}%
}}}}
\put(3751,-4261){\makebox(0,0)[lb]{\smash{{\SetFigFont{8}{9.6}{\rmdefault}{\mddefault}{\updefault}{\color[rgb]{0,0,0}$v_1^1$}%
}}}}
\put(4876,-4261){\makebox(0,0)[lb]{\smash{{\SetFigFont{8}{9.6}{\rmdefault}{\mddefault}{\updefault}{\color[rgb]{0,0,0}$v_2^1$}%
}}}}
\end{picture}%

%% file: sec3.tex
\section{Asymptotic analysis}\label{sec_asymp}
This section intends to illustrate the interest of two-bit decoders over one-bit decoders, in terms of decoding thresholds. In particular, we show that the two-bit decoder, for which expansion conditions for weight-three-error correction has been derived, has better thresholds than one-bit decoders, for various code rates.
\vspace*{-0.5cm}
\subsection{Density evolution}
%\begin{figure*}[!hbt]
\vspace*{-1cm}
{\small
\begin{eqnarray}
P\{W_j=X\}&=&\sum\limits_{\substack{r\in\{-C,C\},n(W),n(S),n(-W):\\f(T,r)=X}}K_{\gamma}P\{R=r\}\prod\limits_{\substack{Y\in M\backslash\{-S\}}}P\{\overline{W}_{j-1}=Y\}^{n(Y)}P\{\overline{W}_{j-1}=-S\}^{n(-S)}\label{DE1}\\
P\{\overline{W}_j=X\}&=&\sum\limits_{\substack{n(W),n(S),n(-W):\\g(n(-S),n(-W),n(W))=X}}K_{\rho}\prod\limits_{\substack{Y\in M\backslash\{-S\}}}P\{W_j=Y\}^{n(Y)}P\{W_j=-S\}^{n(-S)}\label{DE2}
\end{eqnarray}}
%\end{figure*}

Asymptotically in the codeword length, LDPC codes exhibit a threshold phenomenon \cite{Urbanke2001}. In other words, for $\alpha$ smaller than a certain threshold, it is possible to achieve an arbitrarily small bit error probability under iterative decoding, as the codeword length tends to infinity. On the contrary, for noise level larger than the threshold, the bit error probability is always larger than a strictly positive constant, for any codeword length \cite{Urbanke2001,P8}.

In \cite{P8} and \cite{Urbanke2001}, Richardson and Urbanke presented a general method for predicting asymptotic performance of binary LDPC codes. They proved a so-called concentration theorem \cite{P8} according to which decoding performance over any random graph converges, as the code length tends to infinity, to the performance when the graph is cycle-free. Thus, relevant evaluation of performance of binary LDPC codes is possible in the limit case of infinite codeword lengths. The proposed density-evolution method consists in following the evolution of probability densities of messages along the decoding iterations. The messages in each direction are assumed to be independent and identically distributed.

For the class of two-bit decoders, we derive thresholds for different values of $C$ and $S$. The code is assumed to be regular with column weight $\gamma$ and row degree $\rho$. The numbers of $W$, $S$ and $-W$ messages are denoted by $n(W)$, $n(S)$ and $n(-W)$, respectively. In the sets of equations (\ref{DE1}) and (\ref{DE2}), $n(W)\in [0,\dots,d]$, $n(S)\in [0,\dots,d-n(W)]$, $n(-W)\in [0,\dots,d-n(W)-n(S)]$, where $d$ is either $\gamma$ or $\rho$, depending on the context. The number of $-S$ messages $n(-S)$ is hence $d-1-n(W)-n(S)-n(-W)$, with $d=\gamma$ or $\rho$ depending on the context. 
Since the messages of the graph, in each direction, are assumed to be independent and identically distributed, $W_j$ (resp. $\overline{W}_j$) denote the random variables distributed as $\omega_j(v,c)$ (resp. $\varpi_j(c,v)$) for any pair $(v,c)$ of connected variable and check nodes. $X$ denotes an element of the message alphabet $M$. Also, $R\in\{-C,C\}$ denotes the random variable which corresponds to the channem messages. The density evolution equations are given by the sets of equations (\ref{DE1}) and (\ref{DE2}), where:
{\small
\begin{eqnarray}
T&=&\sum_{Y\in M}n(Y)\cdot Y\nonumber\\
K_{\gamma}&=&\binom{\gamma-1}{n(W)}\binom{\gamma-1-n(W)}{n(S)}\binom{\gamma-1-n(W)-n(S)}{n(-W)}\nonumber\\
K_{\rho}&=&\binom{\rho-1}{n(W)}\binom{\rho-1-n(W)}{n(S)}\binom{\rho-1-n(W)-n(S)}{n(-W)}\nonumber
\end{eqnarray}
}
The two functions $f$ and $g$ are defined as follows:
{\small
\begin{eqnarray}
f: \mathbb{Z}^2&\rightarrow& M\nonumber\\
f(T,r) &=& 
\left\{
\begin{array}{lr}
W\cdot sign(T+r),&\mbox{if } 0<|T+r|<S\\
 & \\
S\cdot sign(T+r),&\mbox{if } |T+r|\geq S\\
 & \\
W\cdot sign(r),&\mbox{if } T+r=0
\end{array}
\right.\nonumber\\
&&\nonumber\\
g: \mathbb{N}^3&\rightarrow& M\nonumber\\
g(n_1,n_2,n_3) &=& \nonumber
\end{eqnarray}
\[\left\{
\begin{array}{lr}
W,&\mbox{if } n_3+n_2>0, n_2+n_1=0\bmod (2)\\
 & \\
S,&\mbox{if } n_3+n_2=0, n_2+n_1=0\bmod (2)\\
 & \\
-W,&\mbox{if } n_3+n_2>0, n_2+n_1=1\bmod (2)\\
 & \\
-S,&\mbox{if } n_3+n_2=0, n_2+n_1=1\bmod (2)\\
\end{array}
\right.
\]}

\subsection{Thresholds of two-bit decoders}

\begin{table}[!htb]
\caption{Thresholds of different decoders for column-weight-four codes with row degree $\rho$.}
\label{thresh}
\begin{center}
{\footnotesize
\begin{tabular}{|c|c|c|c|c|c|}
  \hline
  $\rho$ & Rate &  A       & B        &  E       & (1,1,1) \\   \hline
  $8$  & $0.5$  & $0.0474$ & $0.0516$ & $0.0583$ & $0.0467$\\
  $16$ & $0.75$ & $0.0175$ & $0.0175$ & $0.0240$ & $0.0175$ \\
  $32$ & $0.875$& $0.00585$& $0.00585$& $0.00935$& $0.00585$ \\ \hline
  $\rho$ & Rate & $(1,2,1)$& $(1,3,1)$& $(1,4,1)$& $(2,1,1)$\\  \hline
  $8$  & $0.5$  & $0.0509$ & $0.0552$ & $0.0552$ & $0.0467$\\
  $16$ & $0.75$ & $0.0165$ & $0.0175$ & $0.0175$ & $0.0175$\\
  $32$ & $0.875$& $0.00562$& $0.00486$& $0.00486$& $0.00585$\\  \hline
  $\rho$ & Rate &  (2,2,1)  & (2,3,1) & (2,4,1)  & (3,2,1) \\   \hline
  $8$  & $0.5$  &  $0.0567$ & $0.0532$& $0.0552$ & $0.0467$\\
  $16$ & $0.75$ &  $0.0177$ & $0.0168$& $0.0175$ & $0.0218$\\
  $32$ & $0.875$&  $0.00587$& $0.00568$&$0.00486$& $0.00921$\\  \hline
  $\rho$ & Rate & (3,3,1)  & (3,4,1) & (4,3,1)  & (4,4,1)\\   \hline
  $8$  & $0.5$  & $0.0657$ & $0.0620$& $0.0486$ & $0.0657$\\
  $16$ & $0.75$ & $0.0222$ & $0.0203$& $0.0227$ & $0.0222$\\
  $32$ & $0.875$& $0.00755$& $0.00691$&$0.00871$& $0.00755$\\  \hline
  $\rho$ & Rate & Dynamic two-bit & & & \\
   & & decoder with & & &\\
   & & $S=2$ and $W=1$ & & &\\ \hline
  $8$ & $0.5$ & $0.0638$ & & &\\
  $16$ & $0.75$ & $0.0249$ & & &\\
  $32$ & $0.875$ & $0.00953$ & & &\\ \hline
\end{tabular}}
\end{center}
\end{table}

Table \ref{thresh} encompasses thresholds for various code parameters and decoding rules. Thresholds are given in probability of crossover on the BSC. Algorithm E is presented in \cite{P8}. For the two-bit decoders, the set (C,S,W) is given. When the threshold is below $0.001$, $\times$ is put in the box.
The code rate is defined by $1-\frac{\gamma}{\rho}$.
Table \ref{thresh} shows that the specific two-bit decoder with parameters $(C,S,W)=(2,2,1)$, has better thresholds than one-bit decoders Gallager A and B algorithms.
However, this decoder has not the best threshold among the two-bit decoders. Indeed, we tried to achieve a trade-off between good thresholds and not too strong conditions for three error correction. Nevertheless, the method of analysis applied in the proof of the previous section is general, and can be applied to a variety of decoders to obtain similar results.

\textit{Remark:} Algorithm E and the presented dynamic two-bit decoder outperform the other ones, especially for code rates $\frac{3}{4}$ (i.e., $\rho=16$) and $\frac{7}{8}$  (i.e., $\rho=32$). Algorithm E, described in \cite{P8}, is the aforementioned decoder with erasures in the message alphabet. At each iteration, the weight affected to the channel observation (equivalent to $C$ in the two-bit decoder) is optimized \cite{P8}. The dynamic two-bit decoder is based on the same idea: for $S=2$ and $W=1$, $C$ is chosen at each iteration. The better thresholds of the presented dynamic two-bit decoder over Algorithm E indicates that it is interesting to consider decoding on a higher number of bits, even if the channel observation is still one bit, to get better thresholds.

%% file: sec5.tex
\section{Numerical results}

\begin{figure}[!ht]
	\begin{center}
	\includegraphics[width=0.5\textwidth]{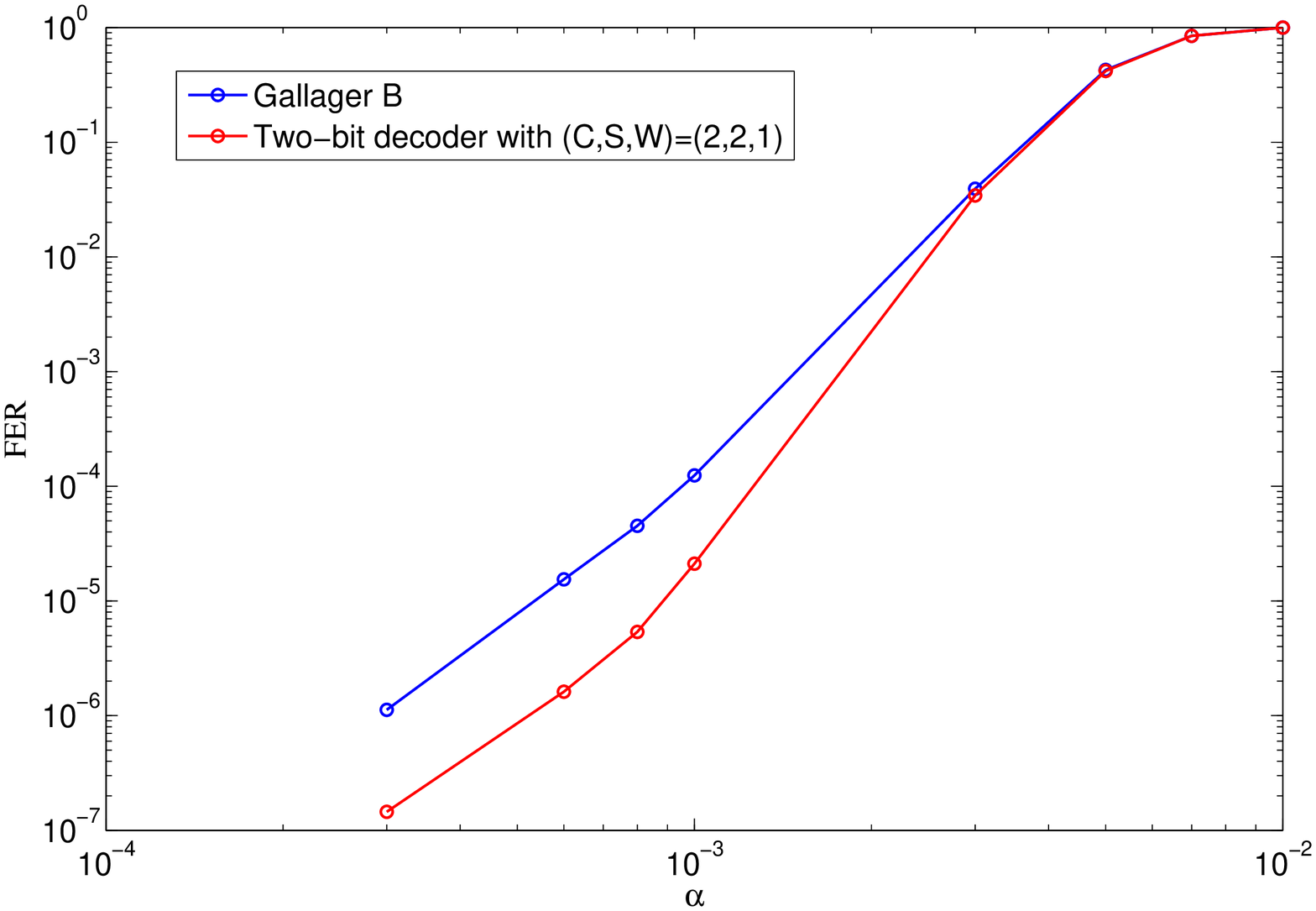}
	\end{center}
\caption{FER versus the crossover probability $\alpha$ for regular column-weight-four MacKay code. The code rate is $0.89$ and the code length is $n=1998$.}
\label{FERquant}
\end{figure}

We have formally proved the capability of weight-three-error correction of an LDPC code satisfying conditions of Theorem \ref{th} and decoded with the two-bit decoder with $(C,S,W)=(2,2,1)$. To compare this two-bit decoder with another one-bit decoder, namely Gallager B, we have plotted FER in Figure \ref{FERquant}. We consider a MacKay code, with column weight four, 1998 variable nodes and 999 check nodes. The code rate is $0.89$. This code has been decoded with Gallager B and the above two-bit decoder. Figure \ref{FERquant} shows that the two-bit decoder has lower FER than Gallager B decoder. In particular, we observe better waterfall performance using the two-bit decoder, and about 1dB gain in the error-floor.

%% file: sec6.tex
\section{Conclusion}

In this paper, we proposed a class of two-bit decoders. We have focused on a specific two-bit decoder for which we have derived necessary and sufficient conditions for a code with Tanner graph of girth six to correct any three errors within three iterations. These conditions are weaker than the conditions for a code to correct three errors when it is decoded with Gallager B algorithm, which uses only one bit. %Hence, two-bit decoder may allow to ensure weight-three error correction capability for higher rate codes than one-bit Gallager type decoding.%We have then given criteria for construction of a code which satisfies the above conditions. 
We have computed thresholds for various two-bit decoders, and shown that the decoder for which the previous conditions has been derived has better thresholds than one-bit decoders, like Gallager A and B. Finally, we have compared the frame error rate performance of the two-bit decoder and Gallager B algorithm for decoding a column-weight-four code with high rate. The two-bit decoder performs better than Gallager B both in the waterfall and in the error-floor region. This illustrates that it is interesting to use two bits rather than one bit for decoding.

Future work includes extending the analysis to derive sufficient conditions to guarantee correction of higher number of errors, as well as investigating on general expression of conditions in terms of the number of quantization bits for the messages. While the focus of the paper was on left-regular LDPC codes, the general methodology and the approach presented in the paper are applicable to irregular LDPC codes also. The analysis in the case of irregular codes will be more complex, but given that carefully designed irregular LDPC codes outperform their regular counterparts, the study of guaranteed error correction under different decoders for such codes is a problem worth investigating.

%% file: Ack.tex
\section*{Acknowledgment}
{\small This work has been done while L. Sassatelli was with ETIS lab, and funded by the French Armament Procurement Agency (DGA).
B. Vasic and S. K. Chilappagari would like to acknowledge the financial support of the NSF (Grants CCF-0634969 and IHCS-0725405).}

%% file: AppendixA.tex
\appendix
In this Appendix, we provide the proofs for Cases 1, 3, 4 and 5 as well as prove the necessity of the conditions stated in Theorem 1.

\textbf{Case 1}: Consider the error configuration shown in Figure \ref{errors}(a). In this case, variables 1, 2 and 3 send incorrect $-W$ messages to their neighbors in the first iteration. They receive $W$ messages from all their neighboring check nodes, they are therefore decoded correctly at the end of first iteration. Error occurs only if there exists a variable node with correct received value that receives four $-W$ messages from its neighboring check nodes (see Table \ref{tabdec}). However, since variables 1, 2 and 3 are the only variables that send incorrect messages in the first iteration, it is impossible to encounter such a variable node without introducing a 4-cycle. Hence, this configuration converges to the correct codeword at the end of the first iteration.\\

\textbf{Case 2}:  Consider the error configuration shown in Figure \ref{errors}(b).\\
In the second half of the first iteration, we have:
{\small
\begin{eqnarray}
\varpi_1(c_4^1,v)&=&-W,\quad v\in \{v_1^1,v_2^1\}\nonumber\\
\varpi_1(c,v)&=&-W,\quad v\in V^2,\quad c\in C^1\backslash c_4^1\nonumber\\
\varpi_1(c,v)&=&W,\quad \mbox{otherwise}\nonumber
\end{eqnarray}}
In the first half of the second iteration, according to Table \ref{tabmp} no $-S$ messages can be sent by variables neither in $V\backslash V^1$ because no $-S$ message propagate in the first iteration, nor variables in $V^1$	 because they all receive at least three $W$ messages:
{\small
\begin{eqnarray}
\omega_2(v,c)&=&-W,\quad v\in \{v_1^1,v_2^1\},\quad c\in C^1\backslash c_4^1\nonumber\\
\omega_2(v,c_4^1)&=&W,\quad v\in \{v_1^1,v_2^1\}\nonumber\\
\omega_2(v_3^1,c)&=&W,\quad c\in C^1\nonumber\\
\omega_2(v,c)&=&-W,\quad v\in \mathcal{N}_0(T_3^1),\quad c \in C^2\nonumber\\
\omega_2(v,c)&=&W,\quad v\in\mathcal{N}_0(T_2^2),\quad c \in C^2\nonumber\\
\omega_2(v,c)&=&W,\quad v\in\mathcal{N}_0(T_3^1),\quad c\in C^1\nonumber\\
\omega_2(v,c)&=&S,\quad \mbox{otherwise}\nonumber
\end{eqnarray}}
%Let us prove that any two variable nodes of type 3-1 cannot share the same check node in $C^2$, in order to ensure that there cannot exist a variable node in $V^2$ which receives four wrong messages at the end of the second iteration.
%\begin{lemma}\label{lem1}
%Any two variable nodes of type 3-1 cannot share the same check node in $C^2$
%\end{lemma}
%\textit{Proof}: The proof is by contradiction. Let $v_1^2, v_2^2 \in V^2$ be variable nodes of type 3-1. Let their common check in $C^2$ be $c^2_1$. Since $v_3^1$ can share at most two checks with $v_1^2, v_2^2$, assume that $c_{10}^1$ and $c_{11}^1$ are not neighbors of $v_1^2$ and $v_2^2$. The neighbors of the variable nodes in the set $\{v_1^1,v_2^1,v_1^2, v_2^2\}$ all belong to the set $\{c^1_1,\ldots,c^1_9\} \cup \{c^2_1\}$ which has cardinality $10$, thus violating the $4 \rightarrow 11$ condition.
%\begin{flushright}$\blacksquare$\end{flushright}
%Hence, with Lemma \ref{lem1}, the messages along the edges in the second half of the second iteration are such that:
In the second half of the second iteration, the messages going out of certain check nodes depend on the connection degree of these check nodes. However, we do not want that the proof be dependent on the degree of connection of check nodes. Hence, we consider in the following the ``worst'' case, that is the configuration where each message has the smallest possible value.
In that case, the messages along the edges in the second half of the second iteration are such that:
{\small
\begin{eqnarray}
\varpi_2(c,v)&=&-W,\quad v\in V^2\cap\mathcal{N}_2(\{v_1^1,v_2^1\}),\quad c\in C^1\backslash c_4^1\nonumber\\
\varpi_2(c_4^1,:)&=&W\nonumber\\
\varpi_2(c,:\backslash v)&=&-W,\quad v\in\mathcal{N}_0(T_3^1),\quad c\in C^2\cap \mathcal{N}_1(T_3^1)\nonumber\\
\varpi_2(c,v)&=&W,\quad v\in V^2,\quad c\in \{c_8^1,c_9^1,c_{S}^1,c_{-S}^1\}\nonumber\\
\varpi_2(c,:)&=&W,\quad c\in C^1\cap \mathcal{N}_1(T_3^1)\nonumber\\
\varpi_2(c,:)&=&W,\quad c\in C^2\cap \mathcal{N}_1(T_2^2)\nonumber\\
\varpi_2(c,v)&=&S,\quad \mbox{otherwise}\nonumber
\end{eqnarray}}
At the end of the second iteration, all $v \in V^1$ receive all correct messages $W$ or $S$. According to Table \ref{tabdec}, all variables in $V^1$ are hence corrected at the end of the second iteration. For variables in $V^2$, since no $-S$ messages propagate in the second half of the second iteration, we see on Table \ref{tabdec} that variables in $V^2$, which are not received in error, are decoded as 1 if and only if they receive four $-W$ messages. The following lemma prove that this is not possible.

\begin{lemma}
No variable node receives four incorrect $-W$ messages at the end of second iteration.
\end{lemma}
\textit{Proof}: 
Let $v$ be such a variable. Then the four neighboring checks of $v$ must belong to $\{c_1^1,c_2^1,c_3^1,c_5^1,c_6^1,c_7^1\}\cup\left(C^2\cap\mathcal{N}_1(T_3^1)\right)$. Note that only two neighbors of $v$ can belong to $\{c_1^1,c_2^1,c_3^1,c_5^1,c_6^1,c_7^1\}$ without introducing a 4-cycle. This implies that there are only three cases: 

\begin{itemize}
\item $v$ has two neighboring checks, say $c_1^2$ and $c_2^2$, in $C^2\cap\mathcal{N}_1(T_3^1)$, and two checks in $\{c_1^1,c_2^1,c_3^1,c_5^1,c_6^1,c_7^1\}$. Let $v_1^2$ and $v_2^2$ be the $T_3^1$ variables connected to $c_1^2$ and $c_2^2$. It results that the set of variables $\{v_1^1,v_2^1,v_1^2,v_2^2,v\}$ is connected to only $11$ checks, which contradicts the 5$\rightarrow$12 expansion condition. This case is hence not possible.
\item $v$ has one neighbor in $\{c_1^1,c_2^1,c_3^1,c_5^1,c_6^1,c_7^1\}$ and three neighbors in $C^2\cap\mathcal{N}_1(T_3^1)$, say $c_1^2$, $c_2^2$ and $c_3^2$. Let $v_1^2$, $v_2^2$ and $v_3^2$ be the $T_3^1$ variables connected to $c_1^2$, $c_2^2$ and $c_3^1$. It results that the set of variables $\{v_1^1,v_2^1,v_1^2,v_2^2,v_3^2,v\}$ is connected to only $13$ checks, which contradicts the 6$\rightarrow$14 expansion condition. This case is hence not possible.
\item $v$ has four neighbors in $C^2\cap\mathcal{N}_1(T_3^1)$, say $c_1^2$, $c_2^2$, $c_3^2$ and $c_4^2$. Let $v_1^2$, $v_2^2$, $v_3^2$ and $v_4^2$ be the $T_3^1$ variables connected to $c_1^2$, $c_2^2$, $c_3^1$ and $c_4^1$. It results that the set of variables $\{v_1^1,v_2^1,v_3^1,v_1^2,v_2^2,v_3^2,v_4^2,v\}$ is connected to only $15$ checks, which contradicts the 8$\rightarrow$16 expansion condition. This case is hence not possible.
\end{itemize}
\begin{flushright}$\blacksquare$\end{flushright}
Hence, the decoder converges at the end of the second iteration.\\

\textbf{Case 3}: Consider the error configuration shown in Figure \ref{errors}(c). In the first iteration, the variables 1, 2 and 3 send incorrect $-W$ messages to their neighboring checks. At the end of the first iteration, they receive correct messages from all their neighboring checks. There is no variable that receives four incorrect messages (as it will cause a four-cycle). Hence, the decoder successfully corrects the three errors.\\

\textbf{Case 5}:
%\begin{figure}
%\begin{center}
%\input{fig3.pstex_t}
%\end{center}
%\caption{Errors configuration for Case 5.}
%\label{cas5}
%\end{figure}
Consider the error configuration shown in Figure \ref{errors}(e).\\
Neither $T_3^1$ nor $T_4^0$ variable can exist in $V^2$ because it would contradict the $4 \rightarrow 11$ expansion condition. At the end of the first iteration, we have:
{\small
\begin{eqnarray}
\varpi_1(c,V^1)&=&W,\quad c\in C^1\backslash\{c_1^1,c_4^1,c_7^1\} \nonumber\\
\varpi_1(c,V^1)&=&-W,\quad c\in \{c_1^1,c_4^1,c_7^1\} \nonumber\\
\varpi_1(c,:\backslash V^1)&=&W,\quad c\in \{c_1^1,c_4^1,c_7^1\} \nonumber\\
\varpi_1(c,:\backslash V^1)&=&-W,\quad c\in C^1\backslash\{c_1^1,c_4^1,c_7^1\} \nonumber\\
\varpi_1(c,:)&=&W,\quad \mbox{otherwise}\nonumber
\end{eqnarray}}
Since a variable in $V^2$ has at most two connections to $C^1$, these variables send $S$ messages to check nodes in $C^1$ at the begining of the second iteration. Hence:
{\small
\begin{eqnarray}
\omega_2(v,c)=-W,\quad v\in V^1,\quad c\in \{c_1^1,c_4^1,c_7^1\} \nonumber\\
\omega_2(v,c)=-S,\quad v\in V^1,\quad c\in C^1\backslash\{c_1^1,c_4^1,c_7^1\} \nonumber\\
\omega_2(v,c)=S,\quad v\in V^2,\quad c\in C^1 \nonumber\\
\omega_2(v,c)=W,\quad v\in \mathcal{N}_0(T_2^2),\quad c\in C\backslash C^1 \nonumber\\
\omega_2(v,c)=S,\quad \mbox{otherwise}\nonumber
\end{eqnarray}}
Hence, at the end of the second iteration, we have:
{\small
\begin{eqnarray}
\varpi_2(c,v)&=&-W,\quad v\in V^1,\quad c\in \{c_1^1,c_4^1,c_7^1\} \nonumber\\
\varpi_2(c,v)&=&S,\quad v\in V^1,\quad c\in C^1\backslash\{c_1^1,c_4^1,c_7^1\} \nonumber\\
\varpi_2(c,v)&=&-S,\quad v\in V\backslash V^1,\quad c\in C^1\backslash\{c_1^1,c_4^1,c_7^1\} \nonumber\\
\varpi_2(c,v)&=&W,\quad v\in V\backslash V^1,\quad c\in \{c_1^1,c_4^1,c_7^1\} \nonumber\\
\varpi_2(c,:)&=&W,\quad c\in \mathcal{N}_1(T_2^2)\backslash C^1 \nonumber\\
\varpi_2(c,:)&=&S,\quad \mbox{otherwise}\nonumber
\end{eqnarray}}
Hence, at the end of the second iteration, a variable in $V^2$ receives only $W$ or $S$ messages from check nodes in $C^2$. It therefore sends $S$ messages to check nodes in $C^1$ at the begining of the third iteration.
As it is used in the sequel, let us mention more explicitly that a pair $(v,c)\in \mathcal{N}_0(T_2^2)\times C^2$ is such that $\mathcal{N}_1((\mathcal{N}_1(v)\cap C^2)\backslash\{c\})\cap \mathcal{N}_0(T_2^2)=\{v\}$ means that the variable $v$ is in $\mathcal{N}_0(T_2^2)$ and the check node which is $(\mathcal{N}_1(v)\cap C^2)\backslash \{c\}$ has no other neighbor in $\mathcal{N}_0(T_2^2)$ except $v$. As well, a pair $(v,c)\in \mathcal{N}_0(T_2^2)\times C^2$ is such that $((\mathcal{N}_1((\mathcal{N}_1(v)\cap C^2)\backslash\{c\}))\cap \mathcal{N}_0(T_2^2))\backslash\{v\}\neq \emptyset$ means that the variable $v$ is in $\mathcal{N}_0(T_2^2)$ and the check node which is $(\mathcal{N}_1(v)\cap C^2)\backslash \{c\}$ has another neighbor in $\mathcal{N}_0(T_2^2)$ different of $v$.
We thus have at the begining of the third iteration:
{\small
\begin{eqnarray}
\omega_3(v,c)&=&W,\quad v\in V^1,\quad c\in \{c_1^1,c_4^1,c_7^1\} \nonumber\\
\omega_3(v,c)&=&-S,\quad v\in V^1,\quad c\in C^1\backslash\{c_1^1,c_4^1,c_7^1\} \nonumber\\
\omega_3(v,c)&=&S,\quad v\in V^2,\quad c\in C^1 \nonumber\\
\omega_3(v,c)&=&W,\quad (v,c)\in \mathcal{N}_0(T_2^2)\times C^2 \nonumber
\end{eqnarray}
\[\mbox{ such that } \mathcal{N}_1((\mathcal{N}_1(v)\cap C^2)\backslash\{c\})\cap \mathcal{N}_0(T_2^2)=\{v\}\]
\begin{eqnarray}
\omega_3(v,c)&=&-W,\quad (v,c)\in \mathcal{N}_0(T_2^2)\times C^2 \nonumber
\end{eqnarray}
\[\mbox{ such that } ((\mathcal{N}_1((\mathcal{N}_1(v)\cap C^2)\backslash\{c\}))\cap \mathcal{N}_0(T_2^2))\backslash\{v\}\neq \emptyset \]
\begin{eqnarray}
\omega_3(v,c)=S,\quad \mbox{otherwise}\nonumber
\end{eqnarray}}
It comes that at the end of the third iteration, for variables in $V^1$ we have:
{\small
\begin{eqnarray}
\varpi_3(c,v)&=&W,\quad v\in V^1,\quad c\in \{c_1^1,c_4^1,c_7^1\} \nonumber\\
\varpi_2(c,v)&=&S,\quad v\in V^1,\quad c\in C^1\backslash\{c_1^1,c_4^1,c_7^1\} \nonumber\\
\end{eqnarray}}
Hence, according to Table \ref{tabdec}, all the variable nodes in $V^1$ are corrected. For messages going into variables not in $V^1$, we have:
{\small
\begin{eqnarray}
\varpi_3(c,v)&=&-S,\quad v\in V^2,\quad c\in C^1\backslash\{c_1^1,c_4^1,c_7^1\} \nonumber\\
\varpi_3(c,v)&=&W,\quad v\in V^2,\quad c\in \{c_1^1,c_4^1,c_7^1\} \nonumber\\
\varpi_3(c,v)&=&W,\quad (v,c) \mbox{ such that } c\in C^2 \nonumber
\end{eqnarray}
\[\mbox{ and there is an even number of }v' \mbox{ in } \mathcal{N}_1(c)\cap\mathcal{N}_0(T_2^2)\backslash\{v\} \mbox{ such that } \]
\[((\mathcal{N}_1((\mathcal{N}_1(v')\cap C^2)\backslash\{c\}))\cap \mathcal{N}_0(T_2^2))\backslash\{v'\}\neq \emptyset\]
\begin{eqnarray}
\varpi_3(c,v)&=&-W,\quad (v,c) \mbox{ such that } c\in C^2 \nonumber
\end{eqnarray}
\[\mbox{ and there is an odd number of }v' \mbox{ in } \mathcal{N}_1(c)\cap\mathcal{N}_0(T_2^2)\backslash\{v\} \mbox{ such that } \]
\[((\mathcal{N}_1((\mathcal{N}_1(v')\cap C^2)\backslash\{c\}))\cap \mathcal{N}_0(T_2^2))\backslash\{v'\}\neq \emptyset\]
\begin{eqnarray}
\varpi_3(c,v)=S,\quad \mbox{otherwise}\nonumber
\end{eqnarray}}
\begin{figure}
\begin{center} \input{cas91.pstex_t} \end{center}
\caption{}
\label{cas91}
\end{figure}
%\begin{figure}
%\begin{center} \includegraphics[width=0.2\textwidth]{cas92.eps} \end{center}
%\caption{}
%\label{cas92}
%\end{figure}
%\begin{figure}
%\begin{center} \includegraphics[width=0.2\textwidth]{cas8.eps} \end{center}
%\caption{}
%\label{cas8}
%\end{figure}
\begin{lemma}
There is no decision error on all variables not in $V^1$ at the end of the third iteration.
\end{lemma}
\textit{Proof}: According to Table \ref{tabdec}, we have to show that the following four situations can not happen:
\begin{itemize}
\item Any variable not in $V^1$ cannot receive more than three $-S$ messages. Indeed, it would imply that the variable has at least three connections to $C^1\backslash\{c_1^1,c_4^1,c_7^1\}$, which would contradict the $4\rightarrow 11$ expansion condition.
\item If any variable, say $v$, not in $V^1$ would receive two $-S$ messages and at least one $-W$ message, it would imply that it has two connections to $C^1\backslash\{c_1^1,c_4^1,c_7^1\}$ and one connection to a check node $c$ such that $c\in C^2$ and there is an odd number of $v'$ in $\mathcal{N}_1(c)\cap\mathcal{N}_0(T_2^2)\backslash\{v\}$ such that $((\mathcal{N}_1((\mathcal{N}_1(v')\cap C^2)\backslash\{c\}))\cap \mathcal{N}_0(T_2^2))\backslash\{v'\}\neq \emptyset$. Let $v''$ denote the variable of such a non-empty set. Then $\{v_1^1,v_2^1,v_3^1,v,v',v''\}$ is connected to only 13 check nodes, contradicting the $6\rightarrow 14$ expansion condition.
\item For sake of clarity, let us now use figures. Without loss of generality, Figures \ref{cas91}(a) and \ref{cas91}(b) illustrate the configurations when the variable $v$ receives, at the end of the third iteration, one $-S$ message and three $-W$ messages, and when it receives one $-S$ message, two $-W$ messages and one $W$ message, respectively. These configurations are not possible as they contradict the $9\rightarrow 18$ expansion condition.
\item Without loss of generality, Figure \ref{cas91}(c) illustrates the configurations when the variable $v$ receives four $-W$ messages at the end of the third iteration. This configuration is not possible as it contradicts the $8\rightarrow 16$ expansion condition.
\end{itemize}
\begin{flushright}$\blacksquare$\end{flushright}
Hence, the decoder converges to the valid codeword at most at the end of the third iteration. This completes the Proof.
\begin{flushright}$\blacksquare$\end{flushright}

\textit{Proof of necessity}:\\
\textbf{Necessity of the $4\rightarrow 11$ condition}\\
Consider the subgraph shown in Figure \ref{cas91}(d). In this case, the $4\rightarrow 11$ condition is not satisfied. It is easy to see that, even though we assume that only $S$ messages are propagated from the check nodes which have an odd degree in the subgraph, to the four variables, the errors are not corrected at the end of the third iteration.
%\begin{figure}
%\begin{center} \includegraphics[width=0.2\textwidth]{A1.eps} \end{center}
%\caption{A $4\rightarrow 10$ subgraph}
%\label{A1}
%\end{figure}

\textbf{Necessity of the $5\rightarrow 12$ condition}\\
As mentioned in \cite{08CKVM}, there exists no graph of girth six which satisfies the $4\rightarrow 11$ condition but does not satisfy the $5\rightarrow 12$ condition.\\

\textbf{Necessity of the $6\rightarrow 14$ condition}\\
Consider the graph shown in Figure \ref{cas91}(e). This graph satisfies the $4\rightarrow 11$ and $5\rightarrow 12$ conditions but not the $6\rightarrow 14$ condition. This graph correspond to the analysis performed above for Case 5. With message values described in this above analysis, it is easy to see that the variables in $V^2$ are wrongly decided to 1 at the end of the third iteration. Hence, in order to guarantee the correction of three errors in three iterations, the $6\rightarrow 14$ condition must be satisfied.\\
%\begin{figure}
%\begin{center} \includegraphics[width=0.2\textwidth]{A2.eps} \end{center}
%\caption{A $6\rightarrow 13$ subgraph}
%\label{A2}
%\end{figure}

\textbf{Necessity of the $8\rightarrow 16$ condition}\\
Consider the graph shown in Figure \ref{cas91}(c). This graph satisfies the $4\rightarrow 11$, $5\rightarrow 12$ and $6\rightarrow 14$ conditions but not the $8\rightarrow 16$ condition. With message values described in the above analysis of Case 5, it is easy to see that $v$ is wrongly decided to 1 at the end of the third iteration. Hence, in order to guarantee the correction of three errors in three iterations, the $8\rightarrow 16$ condition must be satisfied.

\textbf{Necessity of the $9\rightarrow 18$ condition}\\
Consider the graph shown in Figure \ref{cas91}(b). This graph satisfies the $4\rightarrow 11$, $5\rightarrow 12$, $6\rightarrow 14$ and $8\rightarrow 16$ conditions but not the $9\rightarrow 18$. With message values described in the above analysis of Case 5, it is easy to see that the variables in $V^2$ are wrongly decided to 1 at the end of the third iteration. Hence, in order to guarantee the correction of three errors in three iterations, the $9\rightarrow 18$ condition must be satisfied.

\begin{flushright}$\blacksquare$\end{flushright}

%% file: cas91.pstex_t
\begin{picture}(0,0)%
\includegraphics{cas91.pstex}%
\end{picture}%
\setlength{\unitlength}{1579sp}%
\begingroup\makeatletter\ifx\SetFigFont\undefined%
\gdef\SetFigFont#1#2#3#4#5{%
  \reset@font\fontsize{#1}{#2pt}%
  \fontfamily{#3}\fontseries{#4}\fontshape{#5}%
  \selectfont}%
\fi\endgroup%
\begin{picture}(11166,13131)(-397,-13202)
\end{picture}%